\documentclass[12pt]{article}

\usepackage{epsfig,latexsym,amsfonts,amsmath,amsthm,amssymb, mathrsfs,amsbsy,multirow,slashed,wasysym,textcomp,subfigure,hyperref,wrapfig,datetime,comment,mathtools,cancel,cite}
\usepackage[font={footnotesize,sl},bf]{caption}
\usepackage{comment}

\def\be{\begin{equation}}
\def\ee{\end{equation}}
\def\bea{\begin{eqnarray}}
\def\eea{\end{eqnarray}}
 \newcommand{\badat}{\begin{alignedat}}
 \newcommand{\eadat}{\end{alignedat}}

\long\def\new#1\endnew{{\bf #1}}		
\long\def\del#1\enddel{}

\def\del{\partial}

\usepackage{xcolor}
\definecolor{oldmauve}{rgb}{0.4, 0.19, 0.28}
\definecolor{pansypurple}{rgb}{0.47, 0.09, 0.29}
\definecolor{burgundy}{rgb}{0.5, 0.0, 0.13}
\definecolor{carminepink}{rgb}{0.92, 0.3, 0.26}
\definecolor{blue(pigment)}{rgb}{0.2, 0.2, 0.6}
\definecolor{darkseagreen}{rgb}{0.56, 0.74, 0.56}
\definecolor{darkspringgreen}{rgb}{0.09, 0.45, 0.27}
\definecolor{ceruleanblue}{rgb}{0.16, 0.32, 0.75}

\newcommand{\virg}{\hspace{1 mm}, \hspace{8 mm}}

\def\bh{{\bar h}}
\def\bz{{\bar z}}

\def\by{{\bar y}}

\def\bm{{\bar m}}
\def\bJ{{\bar J}}

\begin{document}
\author{Laura Donnay}
\numberwithin{equation}{section} 

\begin{titlepage}
  \thispagestyle{empty}
  
  \begin{center} 
  \vspace*{3cm}
{\LARGE\textbf{MHV leaf amplitudes from parafermions}}

\vskip1cm

   \centerline{Laura Donnay$^{a,b}$\footnote{ldonnay@sissa.it}, Gaston Giribet$^{c}$\footnote{gg1043@nyu.edu}, Beniamino Valsesia$^{a,b}$\footnote{bvalsesi@sissa.it}}
   
\vskip1cm

\it{$^a$SISSA, Via Bonomea 265, 34136 Trieste, Italy}\\
\it{$^b$INFN, Sezione di Trieste, Via Valerio 2, 34127 Trieste, Italy}\\
\it{$^c$Center for Cosmology and Particle Physics\\ Department of Physics, New York University,\\ 726 Broadway, New York City, NY10003, USA.}\\

\end{center}

\vskip1cm

\begin{abstract}
We give a dual CFT representation of MHV leaf amplitudes in the large $N$ and semiclassical limit in terms of non-compact parafermions and a single affine Kac-Moody current for $SO(N)$. This representation is consistent with the other 2D CFT realization of 4D leaf amplitudes proposed in the literature, which is based on a dressed Liouville theory.  The equivalence between the two CFT descriptions arises from the $H_3^+$ WZW-Liouville correspondence, applied to the $SL(2,\mathbb{R})/U(1)$ coset theory in the spectrally flowed sector.
\end{abstract}

\end{titlepage}

\section{Introduction}

Celestial MHV amplitudes can be obtained as a translation invariant combination of the so-called leaf amplitudes, which are amplitudes associated with each 3D hyperbolic ($H_3^+$) foliation of 4D Minkowski spacetime in a Milne expanding patch \cite{Melton:2023bjw}; see also \cite{deBoer:2003vf,Pasterski:2016qvg,Casali:2022fro,Iacobacci:2022yjo,Sleight:2023ojm,Iacobacci:2024nhw,Hao:2023wln,Melton:2023hiq,Melton:2024jyq,Mol:2024etg} for related works. By avoiding the singularities that celestial amplitudes can exhibit, leaf amplitudes provide a framework that facilitates a dual 2D CFT description of celestial correlators. Translation invariance in the bulk indeed typically forces celestial amplitudes to take on a distributional form, which contrasts with the analytic structure one might naturally expect from a standard 2D CFT perspective. This has led to consider the decomposition of celestial amplitudes into contributions from individual leaves of an $H_3^+$ foliation of the bulk spacetime, which are only constrained by conformal invariance. For the case of tree MHV celestial amplitudes, the Kleinian three-point function was shown to be encoded in the residue of a pole in the leaf gluon amplitudes~\cite{Melton:2023bjw}. This example illustrates how the full translation-invariant celestial amplitudes can be reconstituted from non-distributional leaf amplitudes.

 In the recent work \cite{Melton:2024akx}, Melton, Sharma, Strominger and Wang gave a holographic dual description of tree-level 4D MHV scattering amplitudes in terms of a 2D CFT\footnote{See also \cite{Adamo:2022wjo} for the description of the MHV sector of gluon scattering in terms of a 2D celestial CFT using twistor string theory.}. The dual theory, referred to as a `dressed Liouville theory', is a non-compact, non-unitary theory that involves a Liouville\footnote{Connections between celestial amplitudes and Liouville theory were also investigated in \cite{Stieberger:2022zyk, Stieberger:2023fju,Melton:2023lnz,Giribet:2024vnk}.} field together with $N$ free fermions and an additional chiral fermion $\eta$ of negative weight. The leaf amplitudes framework plays a key role in the derivation of a duality between a sector of this theory and MHV celestial amplitudes. It was indeed shown that, in the large $N$ and large central charge $c$ limit, $n$-point correlation functions in the dressed Liouville CFT turn out to be in correspondence with the $n$-point MHV leaf amplitudes on a single $H_3$ slice of Minkowski spacetime. The proof of this statement relies on the fact that, for large $c$, Liouville theory correlation functions can be represented by contact Witten diagrams in Euclidean AdS$_3$ space \cite{Melton:2024akx}. The holographic dictionary is then established in three steps: first, it involves going from the dressed Liouville observables to the so-called Euclidean leaves. Second, going from the Euclidean leaves to the Lorentzian leaves, which involves a continuation from Minkowski to Klein space $\mathbb R^{2,2}$, whose boundary is foliated by celestial tori. Finally, going from the Lorentzian leaves to the full MHV celestial amplitudes, which amounts to single out the pole at $\beta=\sum_{i=1}^n(\Delta_i-1)=0$, where $\Delta_i$ denotes the conformal dimension of the gluons~\cite{Melton:2024akx}.

As already mentioned above, the celestial dual of \cite{Melton:2024akx} consists of a specific non-unitary 2D CFT involving Liouville field theory in direct product with $N$ fermions of conformal weights $(h,\bar h)=(\frac 12, 0)$ and an extra fermion $\eta$ of weights $(-\frac 32, 0)$. A combination of the real (or complex) fermions gives a rank-$N$, level-$1$ affine Kac-Moody ${SO}(N)$ (resp. ${SU}(N)$) current $J^a$ of weights $(1,0)$ which enters in the CFT operator dual to a positive helicity gluon. This current is then dressed with the additional $\eta$ field to construct a $(-1,0)$ current, denoted $\bar J^a$. The latter turns out to be needed in the dual description of bulk fields of negative helicity.

The goal of this paper is to present an alternative 2D CFT realization of 4D MHV leaf amplitudes in terms of a perhaps more familiar theory, namely the Wess-Zumino-Witten (WZW) model. This is achieved by realizing that the dressed Liouville correlators of \cite{Melton:2024akx} can be expressed as specific correlators in the $SL(2,\mathbb{R})/U(1)$ coset theory, which can be realized in terms of parafermions~\cite{Lykken:1988ut, Fateev:1985mm}. Connection between celestial amplitudes and $SL(2,\mathbb{R})$ WZW correlators has recently been explored in the literature \cite{Ogawa:2024nhx, Mol:2024qct,Mol:2024onu,Mol:2024vok}, likely motivated by the fact that, from the leaf amplitudes perspective, it is natural to expect an $H_3^+=SL(2,\mathbb{C})/SU(2)$ structure related to the hyperbolic foliation. However, as we will discuss, a crucial ingredient needed to establish a connection between leaf amplitudes and such WZW correlators is to consider spectrally flowed representations.

The paper is organized as follows. In section \ref{sec:review}, we review the CFT description of MHV leaf amplitudes given in \cite{Melton:2024akx}. In section \ref{sec:new}, we propose an alternative holographic realization of MHV leaf amplitudes in terms of parafermions for the coset theory $SL(2,\mathbb{R})/U(1)$. In appendix \ref{app:rep}, we discuss the relation between the coset relization and the full $SL(2,\mathbb{R})$ WZW theory. Finally, section \ref{sec:OPE} studies the operator product expansion in the 2D theory.

\section{MHV leaf amplitudes and CFT}
\label{sec:review}

This section is a brief review of the 2D CFT realization of MHV leaf amplitudes \cite{Melton:2023bjw, Melton:2024jyq} given in \cite{Melton:2024akx}. It involves a Liouville field theory dressed with extra fields, $N$ free fermions $\psi^i$, and a free chiral fermion $\eta$. Important elements are Liouville primary operators of momentum $\alpha$
\begin{equation}
    V_\alpha(z,\bz)=e^{2\alpha \varphi(z,\bz)}\,,
\end{equation}
which have conformal dimension
\begin{equation}\label{eq:Liouville_h}
    h_{V_{\alpha}}=\bar h_{V_{\alpha}}=\alpha (Q-\alpha)\,.
\end{equation}
The background charge $Q=b+b^{-1}$ is related to the Liouville central charge via $c_L=1+6Q^2$. The dressed operators in \eqref{TYUIO} below will involve `light' Liouville operators, for which the momentum $\alpha$ scales with $b$ in the classical limit $b\to 0$.

The dictionary between bulk positive and negative helicity gluon primaries and dressed \mbox{Liouville} operators is \cite{Melton:2024akx}
\begin{equation}\label{TYUIO}
\badat{2}
    &\mathcal O_{\Delta} ^{+a,\epsilon}(z,\bz)=e^{-i\epsilon \frac{\pi}{2} (\Delta-1)} \lim_{b\to 0}N^+_{\Delta}\, J^a(z) \,V_{\frac b2(\Delta -1)}(z,\bz)\,,\\
    &\mathcal O_{\Delta }^{-a,\epsilon}(z,\bz)=e^{-i\epsilon \frac{\pi}{2} (\Delta+1)} \lim_{b\to 0}N^-_{\Delta  }\, \bJ^a(z) \,V_{\frac b2(\Delta  +1)}(z,\bz)\,,
\eadat
\end{equation}
where the conformal dimension $\Delta$ and the spin $S$ are read from the weights $(h,\bar h)$ using $h=\frac 12 ({\Delta+S})$, $\bar h=\frac 12 ({\Delta-S})$. The index $\epsilon=\pm 1$ labels whether the gluon is outgoing or incoming. The normalization in (\ref{TYUIO}) is
\begin{equation}
\badat{1}
N^+_{\Delta  }&=\mu_{\text{cl}}^{(\Delta  -1)/2}\Gamma(\Delta  -1)e^{-i\pi(\Delta  -1)/2}\,,\\
    N^-_{\Delta  }&=\mu_{\text{cl}}^{\Delta  /2-1/2b^2}e^{\gamma_E-1/b^2}\sqrt{\pi b^3\sin(\pi/b^2)}\Gamma(\Delta  +1)e^{-i\pi(\Delta  +1)/2}\,.
\eadat\label{Anterior}
\end{equation}
The classical Liouville cosmological constant $\mu_{\text{cl}} =\pi\mu b^2$ is held fixed in the classical limit. Importantly, the definition (\ref{TYUIO}) also includes currents $J^a(z)$ and $\bar{J}^a(z)$, for positive helicity and negative gluons, respectively. $J^a(z)$ is built out of the real fermions $\psi^i (z)$ ($i=1,\dots,N$) via
\begin{equation}\label{eq:J}
    J^a(z)=\frac{1}{2}T^a_{ij}:\psi^i\psi^j:(z)
\end{equation}
where $T^a$ are the generators of $so(N)$. From the fermions leading OPE
\begin{equation}
    \psi^i(z)\psi^j(w)=\frac{\delta^{ij}}{z-w}+:\psi^i\psi^j:(w)+\mathcal{O}(z-w)\,,
\end{equation}
one sees that $J^a(z)$ is a Kac-Moody current of weights $(1,0)$ and level $k=1$,
\begin{equation}\label{JJ_ope}
    J^a(z)J^b(w)\sim \frac{\delta^{ab}}{(z-w)^2}+\frac{i{f^{ab}}_cJ^c(w)}{z-w}\,.
\end{equation}
The current $\bar J^a(z)$ in \eqref{TYUIO} is realized by the inclusion of an extra fermionic field $\eta$ of weight $(-\frac 32 ,0)$ as
\begin{equation}
\bJ^a(z)=\eta\partial\eta J^a(z) \,,   
\end{equation}
and is thus of weights $(-1,0)$. 
The $\eta$ field has a regular OPE,
\begin{equation}
    \eta(z)\eta(w)=(z-w)\eta\partial\eta(w)+\mathcal{O}((z-w)^2)\,.
\end{equation}
This operator has four zero modes on the sphere, so to have a non-vanishing correlation function, one has to insert four $\eta$ (see also \cite{Bu:2022dis}) such that
\begin{equation}
    \langle\eta\partial\eta(z)\eta\partial\eta(w)\rangle=(z-w)^4\, .
\end{equation}
From the currents and Liouville weights \eqref{eq:Liouville_h}, one can check that the conformal dimension and spin of dressed operators in \eqref{TYUIO} are given by
\begin{equation}
\badat{2}
   & \Delta_{{\mathcal O}^{+}}=\frac{\Delta  -1}{2}(2+b^2(\Delta  -3))+1 \virg  S_{{\mathcal O}^{+}}=+1\,,\\
   &\Delta_{{\mathcal O}^{-}}=\frac{\Delta  +1}{2}(2+b^2(\Delta  -1))-1 \virg S_{{\mathcal O}^{-}}=-1\,,
\eadat
\end{equation}
which indeed gives, in the classical limit, $\lim_{b \to 0}\Delta_{{\mathcal O}^{\pm }}=\Delta  $. 

To check that correlators of the dressed Liouville CFT operators given in \eqref{TYUIO} reproduce the MHV leaf amplitudes \cite{Melton:2023bjw}, one only needs the following non-vanishing propagators 
\begin{equation}\label{non_vanishing_correlators}
\badat{2}
    &\big \langle \bJ^{a_1}(z_1)\bJ^{a_2}(z_2) \prod_{\ell =3}^n J^{a_{\ell}}(z_{\ell})\big\rangle=\,\text{Tr}(T^{a_1}\dots T^{a_n})\, \frac{z_{12}^4}{z_{12}z_{23}\dots z_{n1}}+\dots\\
    &\big \langle \prod_{\ell =1}^nV_{b \sigma_{\ell }}(z_{\ell },\bar z_{\ell })\big \rangle=\frac{e^{-2\gamma_E+\frac{2}{b^2}}\mu_{\text{cl}}^{\frac{1}{b^{2}}-1-\frac 12 \beta  }}{\pi b^3}\csc[\pi({b^{-2}}- {\beta}/2)]\,\mathcal{C}_{2\sigma_1\dots 2\sigma_n}\,,
    \eadat
\end{equation}
where the dots represent multi-trace terms, $z_{ij}\equiv z_i-z_j$, $\gamma_E$ is the Euler-Mascheroni constant, and
\begin{equation}
    \beta=2\sum_{j}\sigma_j-4\, .
\end{equation}
For finite $b$, $\mathcal{C}_{2\sigma_1\dots 2\sigma_n}$ is a complicated expression, but in the classical limit $b\rightarrow0$ it can be expressed in terms of contact Witten diagrams,
\begin{equation}
    \mathcal{C}_{2\sigma_1\dots 2\sigma_n}=\int_{H_3^+}D^3x \, \prod_{j=1}^nG_{2\sigma_j}(z_j,\bz_j;x)=\int_{H_3^+}\frac{dy\,d\by\, d\rho}{\rho^3} \prod_{j=1}^n\left(\frac{\rho}{\rho^2+|y-z_j|^2}\right)^{2\sigma_j}\,,
\end{equation}
where $D^3x$ is the measure on the unit hyperboloid ($x=\{\rho, y, \bar y \}$) and $G_{2\sigma_j}(z_j,\bz_j;x)$ is the scalar bulk-to-boundary propagator of weight $2\sigma_j$, namely
\begin{equation}
    G_h(z,\bz;x)=\left(\frac{\rho}{\rho^2+|y-z|^2}\right)^h.
\end{equation}
For $N \rightarrow \infty$ we can also simplify the first correlator in \eqref{non_vanishing_correlators} since, in the large $N$ limit, only the first term survives. Suppressing the color indices we can then only focus on color-ordered correlators,
\begin{equation}\label{large_N_current_correlator}
    \langle \bar J(z_1)\bar J(z_2) \prod_{j=3}^n J(z_j)\rangle=\frac{z_{12}^4}{z_{12}z_{23}\dots z_{n1}}.
\end{equation}
Notice that the expression above contains two types of contribution: the term $z_{12}^4$ is given by the $\eta$ bilinear, while the denominator is completely constructed from the dimension-$(1,0)$ currents. Also notice that it has a cyclic structure $z_{12}z_{23}z_{34}z_{45}\dots z_{n1}$ coming from the single trace over $so(N)$ generators. 
Using \eqref{large_N_current_correlator} and \eqref{non_vanishing_correlators}, one thus reproduces~\cite{Melton:2024akx}, for $b\rightarrow0$ and $N\rightarrow\infty$, the MHV leaf amplitude~\cite{Melton:2023bjw}
\begin{equation}
\badat{2}
    &\langle \mathcal{O}_{\Delta_1} ^{-,\epsilon_1}(z_1,\bar z_1)\mathcal{O}_{\Delta_2} ^{-,\epsilon_2}(z_2,\bar z_2) \prod_{j=3}^n \mathcal{O}_{\Delta_j} ^{+,\epsilon_j}(z_j,\bar z_j)\rangle=\prod_{j=1}^n e^{-i\pi \epsilon_j \bar h_j}\, \Gamma(2\bar h_j)\, \frac{z_{12}^4}{z_{12}z_{23}\dots z_{n1}}  \mathcal{C}_{2\bh_1,2\bh_2,\dots,2\bh_n}\,.
    \eadat
\end{equation}
This supports the duality between MHV amplitudes and dressed Liouville CFT correlation functions in the large $N$ limit. 

\section{Leaf amplitudes from parafermions}
\label{sec:new}

In this section, we provide an alternative 2D CFT description of 4D MHV leaf amplitudes in terms of non-compact parafermions~\cite{Lykken:1988ut, Fateev:1985mm}. These parafermions are those that are involved in the realization of the level-$k$ gauged WZW theory on the coset $SL(2,\mathbb{R})/U(1)$. Usually denoted $\Psi^{j}_{m,\bar{m}}(z,\bar{z})$, these parafermions are organized in $j$-representations of $SL(2,\mathbb{R})$ and are closely related to the parafermions considered in the context of $\mathcal{N}=2$ 2D superconformal algebra \cite{Lykken:1988ut}. The two-point functions of such fields have the form
\begin{equation}
\langle \Psi^{j}_{m,\bar m }(z_1,\bar{z}_1)\Psi^{-1-j}_{-m,-\bar m }(z_2,\bar{z}_2)\rangle
=(z_1-z_2)^{\frac{2}{k-2}j(j+1)-\frac 2k m^2}
(\bar z_1-\bar z_2)^{\frac{2}{k-2}j(j+1)-\frac 2k \bar m^2}\,.
\end{equation}
This gives the conformal weights
\begin{equation}
h_{\Psi^j_{m,\bm }}=-\frac{j(j+1)}{k-2}+\frac{m^2}{k}, \ \ \ \ \bar{h}_{\Psi^j_{m,\bm }}=-\frac{j(j+1)}{k-2}+\frac{\bar{m}^2}{k}\,,\label{hpara}
\end{equation}
which realizes the spectrum of the WZW model on the coset $SL(2,\mathbb{R})/U(1)$, see \cite{Witten, DVV, KB, Becker}.

This coset model can also be realized in terms of the $\sigma$-model on $SL(2,\mathbb{R})$ in the Wakimoto free field representation \cite{Wakimoto:1986gf} and supplemented by an extra scalar field $X$ and a $b$-$c$ ghost system that realizes the BRST charge to mode out the $U(1)$ factor. $X$ is charged under the $U(1)$, with the charges being the labels $m, \bar m$; see \cite{KB, DVV, Becker} for details.

Parafermionic $n$-point correlation functions $\langle  \Psi^{j_1}_{m_1,\bar m_1}(z_1,\bz_1)\, ...\, \Psi^{j_n}_{m_n,\bar m_n}(z_n,\bz_n)\rangle$ that obey the specific condition $\sum_{\ell =1}^n(m_\ell+\bar{m}_{\ell})=k(n-2)$ turn out to be proportional to $n$-point correlation functions in Liouville theory; see (\ref{Riba}) below. This is a particular case of a much more general relation that exists between correlation functions in the $SL(2,\mathbb{R})$ WZW model and in Liouville theory, namely \cite{Ribault:2005ms}
\begin{equation}\label{eq:Rib_parafermion}
\badat{2}
    \langle \prod_{\ell=1}^n \Psi^{j_\ell}_{m_\ell,\bar m_\ell}(z_\ell,\bz_\ell) \rangle &= \frac{2\pi^{3}\, b^{1+\frac r2} }{\Gamma (n-1-r)} \prod_{\ell=1}^n \frac{\Gamma(-j_\ell-m_\ell)}{\pi^2\Gamma(j_\ell+1+\bar m_\ell)} \prod_{\ell <\ell'}^nz_{\ell \ell'}^{\beta_{\ell \ell'}}
    \bar{z}_{\ell \ell'}^{\bar{\beta}_{\ell \ell'}}
    \, \int \prod_{a=1}^{n-2-r}d^2 y_a {\prod_{a<a'}|y_{aa'}|^k} \\
    &\quad \times 
    {\prod_{\ell,a}(z_\ell-y_a)^{\frac{k}{2}-m_\ell}}\,
    {(\bar{z}_\ell-\bar{y}_a)^{\frac{k}{2}-\bar{m}_\ell}}
    \, \times\,\langle \prod_{\ell=1}^n  V_{\alpha_\ell} (z_\ell, \bar z_\ell) \prod_{a=1}^{n-2-r} V_{-\frac{1}{2b}} (y_a) \rangle\,,
\eadat
\end{equation}
where 
\begin{equation}\label{eq:beta}
    \beta_{\ell \ell'}=\frac{2}{k}\left(m_\ell-\frac{k}{2}\right)\left(m_{\ell'}-\frac{k}{2}\right), \ \ \ \ \ \  \bar{\beta}_{\ell \ell'}=\frac{2}{k}\left(\bar{m}_\ell-\frac{k}{2}\right)\left(\bar{m}_{\ell'}-\frac{k}{2}\right)
\end{equation}
with the restrictions
\begin{equation}
    \sum_{\ell=1}^n \bm_\ell=\sum_{\ell=1}^n m_\ell=\frac{k}{2}r\virg r\in \mathbb{Z}_{\geq 0}\,.\label{conditione}
\end{equation}
In an unpublished work, Fateev, Zamolodchikov and Zamolodchikov have shown that the bound $r\leq n-2$ also holds, cf. \cite{GN3, Maldacena2001}. We see from \eqref{eq:Rib_parafermion} that, for a violation of the spectral flow condition by $r$ units, the correspondence between WZW and Liouville correlators involves the insertion of $n-2-r$ `degenerate' Liouville operators $V_{-\frac{1}{2b}}$.

While the expectation value on the left hand side of (\ref{eq:Rib_parafermion}) corresponds to the $SL(2,\mathbb{R})/U(1)$ coset theory with central charge $c_{SL(2)/U(1)}=2+\frac{6}{k-2}$, the one on the right hand side corresponds to Liouville theory with central charge $c_L=1+6Q^2$ with $Q=b+\frac 1b$. The dictionary between momenta and parameters in both theories is given by 
\begin{equation}\label{La35}
  b^{2}=\frac{1}{k-2} \virg \alpha_{\ell} = \frac{1}{\sqrt{k-2}}\left(j_{\ell}+\frac k2\right)\,.
\end{equation}

When extended to the full $SL(2,\mathbb{R})$, relation (\ref{eq:Rib_parafermion}) agrees with the analytic extension of the so-called $H_3^+$ WZW-Liouville correspondence\footnote{See appendix \ref{app:rep} for the relation between parafermions and WZW spectral flowed primaries.}. This correspondence was first proven by Ribault and Teschner \cite{Ribault:2005wp} in the spectral flow-preserving case $r=0$,  based on an observed relationship between the Knizhnik-Zamolodchikov equations for WZW models and the Belavin-Polyakov-Zamolodchikov equations for Liouville correlators~\cite{Stoyanovsky}. A generalization of the $H_3^+$ WZW-Liouville correspondence for Riemann surfaces to higher genus $g\geq 0$ in the case $r=-2g$ was given in \cite{Hikida:2007tq} using the path integral approach. Expression (\ref{eq:Rib_parafermion}) for $g=0$ has been proven in \cite{Ribault:2005ms} for the case $0\leq r<n-2$. The case $r=n-2$ was later proven in \cite{Giribet} using the Coulomb gas approach. Here we will be involved with the latter. In that case, we simply have
\begin{equation}
\prod_{\ell=1}^n\hat{N}^{j_{\ell}}_{m_{\ell}, \bar{m}_{\ell}}\,    \, \langle \prod_{\ell=1}^n \Psi^{j_\ell}_{m_\ell,\bar m_\ell}(z_\ell,\bz_\ell) \rangle=\,{2\pi^{3}}\,
    \prod_{\ell <\ell'}^nz_{\ell \ell'}^{\beta_{\ell \ell'}}
    \bar{z}_{\ell \ell'}^{\bar{\beta}_{\ell \ell'}}
    \,\,\, \langle \prod_{\ell=1}^n V_{\alpha_\ell} (z_\ell, \bar z_\ell)\rangle\label{Riba}\,,
\end{equation}
with the normalization
\begin{equation}
\hat{N}^{j_{\ell}}_{m_{\ell}, \bar{m}_{\ell}}=\frac{\pi^2\Gamma(j_\ell+1+\bar m_\ell)}{b^{1/2}\,\Gamma(-j_\ell-m_\ell)}.
\end{equation}

Motivated by the above correspondence between Liouville theory and the spectrally-flowed $H_3^+$ WZW correlators, we identify the CFT operators dual to gluon primaries as,
\begin{equation}\label{parafermion_dictionary}
\badat{2}
    &\, \chi_{2j+k+1}^{+a,\epsilon }(z,\bz)\,=\, e^{-i\pi\epsilon (j+\frac k2)}\,N^{+}_{2j+k+1}\hat{N}^{j}_{+\frac k2, +\frac k2}
    \, J^a(z)\,\Psi^{j}_{+\frac{k}{2},+\frac{k}{2}}(z,\bz)\, ,     \\
&\chi_{2j+k-1,\pm}^{-a,\epsilon }(z,\bz)=  e^{-i\pi\epsilon (j+\frac k2)}\, N^{-}_{2j+k-1}\hat{N}^{j}_{\pm \frac{k'}{2}, \mp \frac{k}{2}}
\,J^a(z)\,\Psi^{j}_{\pm\frac{k'}{2},\mp\frac k2}(z,\bz)\, ,
\eadat
\end{equation}
where $N^{\pm}_{\Delta}$ is given in (\ref{Anterior}) and $k'^{\, 2}=k(k-8)$. As we will argue, in the $k \to \infty$ limit, the positive helicity gluon operator $\mathcal{O}_{\Delta }^{+a,\epsilon }$ in the semiclassical limit corresponds to the parafermion operator $\chi^{+a,\epsilon }$ with $j=-\frac{k}{2}+\frac{1}{2}(\Delta-1)$. The negative helicity gluon $\mathcal{O}_{\Delta }^{-a,\epsilon }$ corresponds to the linear combination $\chi^{-a}=\frac 12(\chi_{+ }^{-a,\epsilon }+\chi_{-}^{-a,\epsilon })$ with $j=-\frac{k}{2}+\frac{1}{2}(\Delta+1)$.

We can already note that amplitudes with $n-1$ operators $\mathcal{O}_{\Delta }^{+a,\epsilon }$ and a single operator $\mathcal{O}_{\Delta }^{-a,\epsilon }$ vanish in virtue of (\ref{conditione}). Let us also mention that the prescription we are providing here is for the mostly plus amplitudes. The expression for the mostly minus amplitudes follow the same line as a $+\leftrightarrow -$ symmetric version of formula (\ref{eq:Rib_parafermion}) exists; this is obtained by changing the $m_{\ell}\to -m_{\ell} $, $\bar m_{\ell}\to -\bar m_{\ell} $ and $r\to -r$ in (\ref{eq:Rib_parafermion})\footnote{This also amounts to change  $\omega_{\ell}\to -\omega_{\ell} $ in the formulae of \cite{Ribault:2005ms}, and the change $+\leftrightarrow -$ makes the $z_{ij}$- and $\bar z_{ij}$-dependence to switch.}. It is worthwhile pointing out that, in contrast with \eqref{TYUIO}, the dressing in the right-hand side of \eqref{parafermion_dictionary} involves the same chiral current $J^a(z)$ for both positive and negative helicity, and hence does not require the introduction of an extra $\eta$ field. 

The conformal weights of the parafermion operators in \eqref{parafermion_dictionary} can be read from (\ref{hpara}) and the values $(h,\bar h )=(1,0)$ for the current $J^a$. We have
\begin{eqnarray}\label{CWs}
\badat{2}
&h_{\chi^{\pm }}=-\frac{j(j+1)}{k-2}+\frac{k}{4}\pm 1\,, \\
&\bar{h}_{\chi^{\pm }}=-\frac{j(j+1)}{k-2}+\frac{k}{4}\,,
\eadat
\end{eqnarray}
so that the spin and the scaling dimension, in the large $k$ limit, are given by
\begin{eqnarray}\label{CW}
S_{\chi^{\pm}}= h_{\chi^{\pm }}-\bar{h}_{\chi^{\pm }}=\pm 1 \virg \lim_{k\to \infty}\Delta_{\chi^{\pm}}=\lim_{k\to \infty} ( h_{\chi^{\pm }}+\bar{h}_{\chi^{\pm }})=\Delta\,.
\end{eqnarray}

Let us now show that correlators of parafermion operators \eqref{parafermion_dictionary} reconstruct MHV leaf amplitudes. The $n$-point function 
\begin{equation}\label{eq:mhvpara}
\mathcal{A}_{j_1, j_2, ...,j_n}\,=\, \frac{1}{2\pi^3}\, \, \langle \chi^{-,\epsilon_1}_{2j_1+k -1}(z_1,\bar{z}_1)\chi^{-,\epsilon_2}_{2j_2+k -1}(z_2,\bar{z}_2)\prod_{\ell =3}^{n}\chi^{+,\epsilon_{\ell }}_{2j_{\ell }+k +1}(z_{\ell},\bar{z}_{\ell}) \rangle \,,
\end{equation}
is computed using \eqref{Riba}.
One can easily see that, since
\begin{equation}
\badat{2}
    m_1&=-m_2=\pm\frac{k'}{2},\quad &m_{\ell >2}=\frac{k}{2},\\
    \bm_1&=-\bm_2=\frac{k}{2},                   &\bm_{\ell >2}=\frac{k}{2},
\eadat
\end{equation}
with $k'^{\, 2}=k(k-8)$ and using \eqref{eq:beta}, the $\bar{z}_{\ell}$-dependent factor disappears from (\ref{Riba}), while the $z_{\ell}$-dependent factor reduces to $z_{12}^4$. This gives precisely the contribution obtained from the $\eta$ bilinear dressing of \cite{Melton:2024akx}. The $n$-point correlator for the $J(z)$ currents, in the large $N$ limit, reproduces the denominator of \eqref{large_N_current_correlator} in the same way while the Liouville correlation function reads,
\be
\langle \prod_{\ell=1}^n V_{\alpha_\ell} (z_\ell, \bar z_\ell)\rangle=\langle \prod_{\ell=1}^n V_{ b\bar h_\ell} (z_\ell, \bar z_\ell)\rangle\,,
\ee
where we used \eqref{La35} and the dictionary below \eqref{parafermion_dictionary}.
Putting everything together, we conclude that, in the large $k$ and large $N$ limit, the $n$-point correlation functions of parafermion  operators (\ref{parafermion_dictionary}) gives
\begin{equation}\label{leaf_amp}
   \mathcal{A}_{j_1, j_2, ...,j_n}=\frac{z_{12}^4}{z_{12}z_{23}\dots z_{n1}}\prod_{\ell=1}^n e^{i\pi \epsilon_\ell \bh_\ell}\Gamma(2\bh_\ell)\mathcal{C}_{2\bh_1\dots 2\bh_n}\,,
\end{equation}
reproducing the large-$N$ limit of the MHV leaf amplitudes. 

We emphasize that the realization of MHV amplitudes in terms of parafermions is totally equivalent to the CFT realization presented in \cite{Melton:2024akx}. The precise relation between both descriptions follows from the generalized version of the $H_3^+$ WZW-Liouville correspondence \cite{Ribault:2005ms, Ribault:2005wp}. The relations between the parameters in the dressed Liouville and parafermion theories are given by $\alpha_{\ell }=b\sigma_{\ell }$ with $\sigma_{\ell}=j_{\ell}+\frac k2$ and $k=2+\frac{1}{b^2}$, and $\sigma_{\ell }=\frac 12 (\Delta_{\ell }\mp 1)$ for operators of helicity $\pm 1$. 
Nevertheless, it is worthwhile emphasizing that the dual description in terms of parafermions only involves a single Kac-Moody current and hence does not involve the addition of an extra $(-\frac32,0)$-weight free fermion in the description of bulk conformal primary gluons of negative helicity. 
In particular, one can see that certain correlators that involve more than two negative helicity parafermion operators~\eqref{parafermion_dictionary} can be non-vanishing. For example, correlators involving four operators $\chi^{-a}$ and $n-4$ operators $\chi^{+a}$ would correspond, in the Liouville theory, to ($n+2$)-point functions involving two degenerate fields $V_{-\frac{1}{2b}}$ and additional field dressing. While do not provide here a bulk interpretation of the correlators involving such operators, it would be very interesting to understand the role played by the insertion of a Liouville degenerate operator from the gauge theory perspective. Similarly, understanding the role of the so-called spectral flow operator in the $SL(2,\mathbb{R})/U(1)$ theory from the gauge theory perspective is an interesting question to explore. Here above, we showed the dual CFT realization in terms of operators (\ref{parafermion_dictionary}), understood as a correspondence between the correlators \eqref{eq:mhvpara} and MHV amplitudes. 
In Appendix \ref{app:rep}, we discuss whether the parafermionic realization for the coset $SL(2,\mathbb{R})/U(1)$ theory can be uplifted to the full level-$k$ $SL(2,\mathbb{R})$ WZW theory.

\section{Operator product expansions}
\label{sec:OPE}
OPE of leaf amplitudes can be obtained from the collinear expansion of higher-point leaf amplitudes~\cite{Melton:2024jyq}; the leading $z_{12}$ pole is given by 
\begin{equation}
\badat{2}\label{Str_OPEs}
    \mathcal{O}_{\Delta_1}^+(z_1,\bz_1)\mathcal{O}_{\Delta_2}^+(z_2,\bz_2)&\sim\frac{1}{z_{12}}B(\Delta_1-1,\Delta_2-1)\mathcal{O}_{\Delta_1+\Delta_2-1}^+(z_2,\bz_2)+\dots\\
    \mathcal{O}_{\Delta_1}^+(z_1,\bz_1)\mathcal{O}_{\Delta_2}^-(z_2,\bz_2)&\sim\frac{1}{z_{12}}B(\Delta_1+1,\Delta_2-1)\mathcal{O}_{\Delta_1+\Delta_2-1}^-(z_2,\bz_2)+\dots\\
\eadat
\end{equation}
where $B(x,y)$ is the Euler beta function and the dots stand for descendants. It is instructive to compute the corresponding OPE from the dual CFT description \eqref{TYUIO} in terms of Liouville theory dressed with affine currents (see also \cite{Melton:2023lnz}). To see this explicitly, consider first the Liouville OPE
\begin{equation}
    V_{\alpha_1}(z_1,\bz_1)V_{\alpha_2}(z_2,\bz_2)=\frac{1}{\pi}\int_{\alpha \in \frac Q2 +i\mathbb{R}} d\alpha\, C(\alpha_1,\alpha_2,Q-\alpha)|z_{12}|^{2(h_\alpha-h_{\alpha_1}-h_{\alpha_2})}V_{\alpha}(z_2,\bz_2)+\dots
\end{equation}
where $C(\alpha_1,\alpha_2,\alpha_3)$ are the DOZZ Liouville structure constants\cite{DO, ZZ}
\begin{equation}
\badat{2}
    C(\alpha_1,\alpha_2,\alpha_3)&=\left[\pi \mu \gamma(b^2)b^{2-2b^2}\right]^{\left(Q-\alpha_{123}\right)/b}\frac{\Upsilon'(0)\Upsilon(2\alpha_1)\Upsilon(2\alpha_2)\Upsilon(2\alpha_3)}{\Upsilon(\alpha_{123}-Q)\Upsilon(\alpha_{12}^3)\Upsilon(\alpha_{13}^2)\Upsilon(\alpha_{23}^1)}\,,
\eadat
\end{equation}
where we have introduced the notation $\alpha_{ijk}=\alpha_i+\alpha_j+\alpha_k,\ \alpha_{ij}^k=\alpha_i+\alpha_j-\alpha_k$ and
\begin{equation}
    \gamma(x)=\frac{\Gamma(x)}{\Gamma(1-x)}\,.
\end{equation}
We refer the reader to \cite{DO, ZZ}  for notations and definitions of the $\Upsilon$-function. These structure constants obey the reflection property
\begin{equation}
\badat{2}
C(\alpha_1,\alpha_2,Q-\alpha_3)=C(\alpha_1,\alpha_2,\alpha_3)\, S\left(\alpha_3-\frac{Q}{2}\right),
\eadat
\end{equation}
with the reflection coefficient being
\begin{equation}
\badat{2}
S(P)=-\left(\pi \mu \gamma(b^2)\right)^{2P/b}\frac{\Gamma(1-2P/b)\Gamma(1-2Pb)}{\Gamma(1+2P/b)\Gamma(1+2Pb)}\, .
\eadat
\end{equation}
We now compute $c(\sigma_1,\sigma_2,\sigma_3)\equiv b\, C(b\sigma_1,b\sigma_2,Q-b\sigma_3)$ in the semiclassical limit ($b\to 0$), which yields
\begin{equation}
\badat{2}
c(\sigma_1,\sigma_2,\sigma_3)=\frac{1}{2}\mu_{\mathrm{cl}}^{-\sigma_{12}^3}\, \frac{\sin\pi(2\sigma_3-1/b^2)}{\sin\pi(\sigma_{123}-1/b^2)}\frac{\Gamma(\sigma_{12}^3)\Gamma(\sigma_{13}^2)\Gamma(\sigma_{23}^1)\Gamma(\sigma_{123}-1)}{\Gamma(2\sigma_1)\Gamma(2\sigma_2)\Gamma(2\sigma_3-1)}
\eadat
\end{equation}
so that the OPE can be written as
\begin{equation}\label{Liouville_soft_OPE}
    V_{b\sigma_1}(z_1,\bz_1)V_{b\sigma_2}(z_2,\bz_2)=\frac{1}{\pi}\int d\sigma\, c(\sigma_1,\sigma_2,\sigma)|z_{12}|^{2(\sigma-\sigma_1-\sigma_2)}V_{b\sigma}(z_2,\bz_2)+\dots
\end{equation}
where we consider the summation over light operators, as heavy operators decouple when $b\to 0$. The ellipsis stands for higher-order contributions. 

Using \eqref{Liouville_soft_OPE}, the currents OPE \eqref{JJ_ope} and the dictionary \eqref{TYUIO} of \cite{Melton:2024akx}, we find
\begin{equation}
\badat{2}\label{OPE_Strominger}
    \mathcal{O}^{a_1,+}_{\Delta_1}(z_1,\bz_1)\mathcal{O}^{a_2,+}_{\Delta_2}(z_2,\bz_2)
    &= \int_{i\mathbb{R}_+} d\sigma\, c\left(\frac{1}{2}(\Delta_1-1),\frac{1}{2}(\Delta_2-1),\sigma\right)e^{-i\pi\left(\Delta_1+\Delta_2-2-2\sigma\right)}\mu^{\frac{1}{2}(\Delta_1+\Delta_2-2-2\sigma)}\\
    &\times|z_{12}|^{2\sigma-\Delta_1-\Delta_2+2}\, \,  \frac{\Gamma(\Delta_1-1)\Gamma(\Delta_2-1)}{2\pi\,\Gamma(2\sigma)}\\
    &\times\left(\frac{\delta^{a_1a_2}}{z_{12}^2}\psi_{2\sigma+1}^+(z_2,\bz_2)+\frac{i{f^{a_1a_2}}_a}{z_{12}}\mathcal{O}^{a,+}_{2\sigma+1}(z_2,\bz_2)\right)+\dots
\eadat
\end{equation}
where
\begin{equation}
    \psi_\Delta^+(z,\bz)=e^{-i\frac{\pi}{2} (\Delta-1)}N_\Delta^+\,V_{\frac{b}{2}(\Delta-1)}(z,\bz)
\end{equation}
is a normalized Liouville field. The contribution to the OPE (\ref{OPE_Strominger}) corresponds to $2\sigma =\Delta_1 +\Delta_2 -2$. In that case one recovers the OPE in \cite{Melton:2023lnz}; see (4.7) therein. Notice that, for the celestial operators to remain on the principal line, we have to constrain the integral over $\sigma$ to pure imaginary values, $\sigma\in i\mathbb{R}_+$. Also, notice that, if $\Delta_{1,2}$ lie on the principal line, the exponent of $|z_{12}|$ in the OPE is purely imaginary. However we notice that for $\sigma=\frac 12 \Delta_1+\frac 12 \Delta_2-1$ this exponent is vanishing and the Liouville structure constant develops a simple pole on the imaginary axis; namely
\begin{equation}
    c\left(\frac{1}{2}(\Delta_1-1),\frac{1}{2}(\Delta_2-1),\sigma\right)\,\sim \,\frac{1}{\Delta_1+\Delta_2-2-2\sigma}
\end{equation}
for $\sigma\simeq \frac 12 \Delta_1+\frac 12 \Delta_2-1$.
By deforming the integration path we pick up the residue and we see that the OPE receives a contribution of the form
\begin{equation}\label{Almost_Leaf}
\badat{2}
    \mathcal{O}^{a_1,+}_{\Delta_1}(z_1,\bz_1)\mathcal{O}^{a_2,+}_{\Delta_2}(z_2,\bz_2)&
    \sim\,  B(\Delta_1-1,\Delta_2-1)\left(\frac{\delta^{a_1a_2}}{z_{12}^2}\psi_{\Delta_1+\Delta_2-1}^+(z_2,\bz_2)+\frac{i{f^{a_1a_2}}_a}{z_{12}}\mathcal{O}^{a,+}_{\Delta_1+\Delta_2-1}(z_2,\bz_2)\right).
\eadat
\end{equation}
The $z^{-1}_{12}$ term precisely coincides with the leaf OPE, cf. \cite{Melton:2024jyq}. In addition, we have the extra term
\begin{equation}
    \frac{\delta^{a_1a_2}}{z_{12}^2}\psi_{\Delta_1+\Delta_2-1}^+(z_2,\bz_2)\,,
\end{equation}
which is a scalar contribution that contains a Liouville operator. The latter comes from the central term in the current algebra. One might also consider a realization of the affine current algebra at level $\kappa $, e.g. coming from a $SO(N)$ WZW factor. This would result in a current OPE
\begin{equation}
    J^{a_1}(z_1)J^{a_2}(z_2)=\frac{\kappa\, \delta^{a_1a_2}}{z_{12}^2}+\frac{i{f^{a_1a_2}}_{a_3}}{z_{12}}J^{a_3}(z_2)+\dots
\end{equation}
{It has been observed in \cite{Melton:2023lnz} that a similar scalar contribution can be interpreted from a bulk perspective as an interaction between gluons and an extra bulk scalar field.}

Let us now move to the OPE produced from the parafermion representation \eqref{parafermion_dictionary}. In the case of positive helicity operators, we obtain
\begin{equation}
\badat{2}
    \Psi^{j_1}_{\frac{k}{2},\frac{k}{2}}(z_1,\bz_1) \Psi^{j_2}_{\frac{k}{2},\frac{k}{2}}(z_2,\bz_2)
    &\sim \frac{1}{\hat N^{j_1}_{\frac{k}{2},\frac{k}{2}}\hat N^{j_2}_{\frac{k}{2},\frac{k}{2}}}\int {dj}\, |z_{12}|^{\Delta-\Delta_1-\Delta_2+1}\, 
    \tilde{C}_{\text{WZW}}(j_1,j_2,j^*)\,{\hat N^{j}_{\frac{k}{2},\frac{k}{2}}}
    \,\Psi^{j}_{\frac{k}{2},\frac{k}{2}}(z_2,\bz_2) 
\eadat
\end{equation}
where $j^*\equiv -j-1$ and where
\begin{equation}
\badat{2}
    \tilde{C}_{\text{WZW}}(j_1,j_2,j_3)\,=\,c_k\, C(\alpha_1,\alpha_2,\alpha_3)\\
\eadat
\end{equation}
are $SL(2,\mathbb{R})$ WZW structure constants in the spectral flow sector with $\omega_1 +\omega_2 -\omega_3  =-1$, with $c_k$ being a $k$-dependent factor. The explicit expression of $\tilde{C}_{\text{WZW}}(j_1,j_2,j_3)$ has been computed in \cite{GN3, Maldacena2001, Giribet,Ribault:2005ms}; see those references for details. Considering \eqref{La35}, it can easily be shown that $\tilde{C}_{\text{WZW}}(j_1,j_2,j^*)$ agree with the Liouville structure constants ${C}(\alpha_1,\alpha_2,Q-\alpha)$, and so the OPE. The conformal dimension $\Delta $ is related to $j$ as in (\ref{CWs})-(\ref{CW}). We still have to prescribe how to integrate in the $j$-plane, which is done by requiring the conformal weights of the celestial theory lying on the principal line. Putting all this together, we obtain
\begin{equation}\label{OPE_GDV}
\badat{2}
    \mathcal{O}^{a_1,+}_{2j_1+k+1}(z_1,\bz_1)\mathcal{O}^{a_2,+}_{2j_2+k+1}(z_2,\bz_2)
    &=\int {dj}\,\tilde{C}_{\text{WZW}} \left(j_1,j_2,j^*\right)|z_{12}|^{\Delta-\Delta_1-\Delta_2+1}\\
    &\times e^{-i\pi(\Delta_1+\Delta_2-\Delta-1)}\mu^{\frac{1}{2}(\Delta_1+\Delta_2-\Delta-1)}\frac{\Gamma(\Delta_1-1)\Gamma(\Delta_2-1)}{{4 b^{1/2}}\,\Gamma(\Delta-1)}\\
    &\times\left(\frac{\delta^{a_1a_2}}{z_{12}^2}\tilde{\psi}_{2j+k+1}^+(z_2,\bz_2)+\frac{i{f^{a_1a_2}}_a}{z_{12}}\mathcal{O}^{a,+}_{2j+k+1}(z_2,\bz_2)\right)+\dots
\eadat
\end{equation}
where the first term contains an analogous scalar contribution
\begin{equation}
    \tilde{\psi}_{2j+k+1}^+(z_2,\bz_2)=e^{-i\pi(j+\frac k2)}\,N^{+}_{2j+k+1}\hat N^{j}_{\frac k2, \frac k2}\, 
    \Psi^{j}_{\frac{k}{2},\frac{k}{2}}(z,\bz)\, .
\end{equation}
In the semiclassical limit, for $\Delta$ to lie on the principal line, we can consider $j=-\frac k2+\frac i2 {\lambda}$, $\lambda\in\mathbb{R}$. Identifying $\sigma=\frac i2 {\lambda}$, we see that \eqref{OPE_GDV} reproduces the same structure as in \eqref{OPE_Strominger}; in the semiclassical limit, we recover the OPE (\ref{Almost_Leaf}). 

In summary, we showed that it is possible to recast the CFT$_2$ construction of the large $N$ MHV leaf amplitudes proposed in \cite{Melton:2024akx} in terms of non-compact parafermions in direct product with $(1,0)$ affine Kac-Moody currents. This formulation does not involve the introduction of additional fields of negative dimension in the dressing of negative helicity gluon operators and follows from the $H_3^+$ WZW-Liouville correspondence  \cite{Stoyanovsky, Ribault:2005ms, Ribault:2005wp, Giribet:2005ix}, applied to the spectrally flowed sector and restricted to the $SL(2,\mathbb{R})/U(1)$ coset theory. 
We hope that the 2D CFT realization of leaf amplitudes in terms of the $H_3^+$ WZW model presented here will help derive new insights for celestial CFTs inspired by the AdS$_3$/CFT$_2$ correspondence.

\subsection*{Acknowledgments}
The authors thank Walker Melton and Andy Strominger for helpful comments. L.D. and B.V. are supported by the European Research Council
(ERC) Project 101076737 - CeleBH. Views and opinions
expressed are however those of the author only and do
not necessarily reflect those of the European Union or
the European Research Council. Neither the European
Union nor the granting authority can be held responsible for them. L.D. and B.V. are also partially supported by INFN
Iniziativa Specifica ST\& FI. L.D.'s research was also supported in part by the Simons Foundation through the
Simons Foundation Emmy Noether Fellows Program at
Perimeter Institute. Research at Perimeter Institute is
supported in part by the Government of Canada through
the Department of Innovation, Science and Economic Development and by the Province of Ontario through the Ministry of Colleges and Universities. 
B.V. thanks the CCPP at New York University for hosting him during the preparation of this work. B.V. was also supported by the European program Erasmus+.

\appendix
\section{$H_3^+$ representation}
\label{app:rep}

Primary fields of the $SL(2,\mathbb{R})$ WZW theory, including those in the so-called spectrally flowed representation, can be expressed in terms of parafermions $\Psi^j_{m,\bar m}$ for the coset model $SL(2,\mathbb{R})/U(1)$ and an extra $U(1)$ field. Denoting by $\Phi^{j,w}_{m,\bar m }(z,\bz)$ fields of isospin $j$ and spectral flow number $w\in \mathbb Z$, the relation is given by~\cite{Maldacena2001, GN2, GN3}
\begin{equation}\label{eq:flowed}
    \Phi^{j,w}_{m,\bm}(z,\bz)\, =\, e^{i\sqrt{\frac 2k}\left(m+w\frac{k}{2}\right)\phi(z)+i\sqrt{\frac 2k}\left(\bar{m}+w\frac{k}{2}\right)\bar\phi (\bar z )}\, \Psi^{j}_{m,\bm}(z,\bar z )\,,
\end{equation}
with the timelike free boson $\phi (z,\bar z )=\phi ( z )+\bar\phi (\bar z )$. The role of this field is to restore the $U(1)$ factor of the coset theory, which is realized by the current $J^3(z)=-i\sqrt{\frac k2}\partial\phi (z)$ obeying the operator product expansion
\begin{equation}
J^3(z)\, \Phi^{j,w}_{m,\bar m }(w,\bar{w}) \,\sim \,\left( {m+\frac k2 w}\right)\, \frac{\Phi^{j,w}_{m,\bar m }(w,\bar{w})}{(z-w)}\,.
\end{equation}
The conformal weights of the spectral flowed fields \eqref{eq:flowed} are
\begin{equation}
\badat{2}
    &h_{\Phi _{m\bar m }^{j,w}}=-j(j+1)b^2-wm-\frac{2+b^{-2}}{4}w^2\, , \\
    &\bar h_{\Phi _{m\bar m }^{j,w}}=-j(j+1)b^2-w\bar m-\frac{2+b^{-2}}{4}w^2\,,
    \eadat
\end{equation}
with $b^{-2}=k-2$.

$H_3^+$ WZW-Liouville correspondence~\cite{Ribault:2005ms} leads to write the correlators $\langle \prod_{\ell=1}^n \Phi^{j_\ell,w_\ell}_{m_\ell,\bar m_\ell}(z_\ell,\bz_\ell) \rangle$ in terms of Liouville correlation functions $\langle \prod_{\ell=1}^n V_{\alpha_\ell} (z_\ell, \bar z_\ell) \prod_{a=1}^{n-2-r} V_{-\frac{1}{2b}} (y_a) \rangle$ through a formula that looks like (\ref{eq:Rib_parafermion}), but replacing
\begin{eqnarray}\label{wzw_betas}
&&\beta_{\ell \ell'}\to \beta_{\ell \ell'}-\frac 2k\left(m_{\ell}+\frac k2 w_{\ell}\right) \left(m_{\ell '}+\frac k2 w_{\ell '}\right)\nonumber \\
&&\bar\beta_{\ell \ell'}\to \bar \beta_{\ell \ell'}-\frac 2k\left(\bar m_{\ell}+\frac k2 w_{\ell}\right) \left(\bar m_{\ell '}+\frac k2 w_{\ell '}\right)\,,
\end{eqnarray}
together with the restrictions
\begin{eqnarray}\label{WZW_cond}
 \sum_{\ell =1}^n m_\ell=\sum_{\ell =1}^n \bar{m}_\ell=\frac{k}{2}r \virg \sum_{\ell=1}^n w_\ell=-r\,.
\end{eqnarray}
The above conditions ensure the conservation of the charge associated to the $U(1)$ current, namely
\begin{equation}
    \sum _{\ell =1}^n\left(m_{\ell}+\frac k2 w_{\ell} \right) = \sum _{\ell =1}^n\left(\bar{m}_{\ell}+\frac k2 w_{\ell} \right) = 0 
    \,.
\end{equation}

Using the above relation between Liouville theory and the WZW model, it would be natural to directly propose a map between gluon operators and $H_3$ WZW fields of the form
\begin{equation}\label{proposed_WZW_mapping}
    \mathcal O_{\Delta} ^{\pm \, a,\, \epsilon }(z,\bar{z})\, \propto J^a(z)\,\Phi^{j,w_{\pm}}_{m,\bar m}(z,\bar{z})\,,
\end{equation}
for some specific value of $j,w_{\pm},m$ and $\bar m$. To see if this is possible, let us have a closer look at what terms in \eqref{leaf_amp} can be reproduced from an $n$-point function of operators of the form \eqref{proposed_WZW_mapping}. From \eqref{leaf_amp}, it is clear that the cyclic $1/(z_{12}z_{23}\dots z_{n1})$ factor will come again from the current $n$-point function. This means that the $SL(2,\mathbb{R})$ WZW amplitude should be able to reproduce both the Liouville and the $z_{12}^4$ contribution. However, it turns out that, due to the constraints specified in \eqref{WZW_cond} and the values of the $\beta_{\ell\ell'}$ exponents given in \eqref{wzw_betas}, one can show that no combination of $j, w_{\pm}, m, \bar{m}$ can result in a consistent dictionary that reproduces the factor $z_{12}^4$. This indicates that constructing gluon operators out of a $SL(2,\mathbb{R})$ primary and a Kac-Moody current would necessarily require the introduction of additional fields.

\bibliographystyle{style}
\bibliography{references}

\providecommand{\href}[2]{#2}\begingroup\raggedright\begin{thebibliography}{10}

\bibitem{Melton:2023bjw}
W.~Melton, A.~Sharma  and A.~Strominger, \emph{{Celestial leaf amplitudes}}, JHEP {\bf 07} (2024) 132,
\href{http://www.arXiv.org/abs/2312.07820}{{\tt 2312.07820}}

\bibitem{deBoer:2003vf}
J.~de~Boer and S.~N. Solodukhin, \emph{{A Holographic reduction of Minkowski space-time}}, Nucl. Phys. {\bf B665} (2003) 545--593,
\href{http://www.arXiv.org/abs/hep-th/0303006}{{\tt hep-th/0303006}}

\bibitem{Pasterski:2016qvg}
S.~Pasterski, S.-H. Shao  and A.~Strominger, \emph{{Flat Space Amplitudes and Conformal Symmetry of the Celestial Sphere}}, Phys. Rev. {\bf D96} (2017), no.~6, 065026,
\href{http://www.arXiv.org/abs/1701.00049}{{\tt 1701.00049}}

\bibitem{Casali:2022fro}
E.~Casali, W.~Melton  and A.~Strominger, \emph{{Celestial amplitudes as AdS-Witten diagrams}}, JHEP {\bf 11} (2022) 140,
\href{http://www.arXiv.org/abs/2204.10249}{{\tt 2204.10249}}

\bibitem{Iacobacci:2022yjo}
L.~Iacobacci, C.~Sleight  and M.~Taronna, \emph{{From Celestial Correlators to AdS, and back}}, JHEP {\bf 06} (2022) 053,
\href{http://www.arXiv.org/abs/2208.01629}{{\tt 2208.01629}}

\bibitem{Sleight:2023ojm}
C.~Sleight and M.~Taronna, \emph{{Celestial Holography Revisited}}, Phys. Rev. Lett. {\bf 133} (2024), no.~24, 241601,
\href{http://www.arXiv.org/abs/2301.01810}{{\tt 2301.01810}}

\bibitem{Iacobacci:2024nhw}
L.~Iacobacci, C.~Sleight  and M.~Taronna, \emph{{Celestial holography revisited. Part II. Correlators and K\"all\'en-Lehmann}}, JHEP {\bf 08} (2024) 033,
\href{http://www.arXiv.org/abs/2401.16591}{{\tt 2401.16591}}

\bibitem{Hao:2023wln}
Z.~Hao and M.~Taylor, \emph{{Flat holography and celestial shockwaves}}, JHEP {\bf 02} (2024) 090,
\href{http://www.arXiv.org/abs/2309.04307}{{\tt 2309.04307}}

\bibitem{Melton:2023hiq}
W.~Melton, A.~Sharma  and A.~Strominger, \emph{{Conformal correlators on the Lorentzian torus}}, Phys. Rev. D {\bf 109} (2024), no.~10, L101701,
\href{http://www.arXiv.org/abs/2310.15104}{{\tt 2310.15104}}

\bibitem{Melton:2024jyq}
W.~Melton, A.~Sharma  and A.~Strominger, \emph{{Soft algebras for leaf amplitudes}}, JHEP {\bf 07} (2024) 070,
\href{http://www.arXiv.org/abs/2402.04150}{{\tt 2402.04150}}

\bibitem{Mol:2024etg}
I.~Mol, \emph{{A Holographic Construction of MHV Graviton Amplitudes in Celestial CFT}},
\href{http://www.arXiv.org/abs/2408.10944}{{\tt 2408.10944}}

\bibitem{Melton:2024akx}
W.~Melton, A.~Sharma, A.~Strominger  and T.~Wang, \emph{{Celestial Dual for Maximal Helicity Violating Amplitudes}}, Phys. Rev. Lett. {\bf 133} (2024), no.~9, 091603,
\href{http://www.arXiv.org/abs/2403.18896}{{\tt 2403.18896}}

\bibitem{Adamo:2022wjo}
T.~Adamo, W.~Bu, E.~Casali  and A.~Sharma, \emph{{All-order celestial OPE in the MHV sector}}, JHEP {\bf 03} (11, 2022) 252,
\href{http://www.arXiv.org/abs/2211.17124}{{\tt 2211.17124}}

\bibitem{Stieberger:2022zyk}
S.~Stieberger, T.~R. Taylor  and B.~Zhu, \emph{{Celestial Liouville theory for Yang-Mills amplitudes}}, Phys. Lett. B {\bf 836} (2023) 137588,
\href{http://www.arXiv.org/abs/2209.02724}{{\tt 2209.02724}}

\bibitem{Stieberger:2023fju}
S.~Stieberger, T.~R. Taylor  and B.~Zhu, \emph{{Yang-Mills as a Liouville theory}}, Phys. Lett. B {\bf 846} (2023) 138229,
\href{http://www.arXiv.org/abs/2308.09741}{{\tt 2308.09741}}

\bibitem{Melton:2023lnz}
W.~Melton and S.~A. Narayanan, \emph{{Celestial gluon amplitudes from the outside in}}, JHEP {\bf 05} (2024) 211,
\href{http://www.arXiv.org/abs/2312.12394}{{\tt 2312.12394}}

\bibitem{Giribet:2024vnk}
G.~Giribet, \emph{{Remarks on celestial amplitudes and Liouville theory}},
\href{http://www.arXiv.org/abs/2403.03374}{{\tt 2403.03374}}

\bibitem{Lykken:1988ut}
J.~D. Lykken, \emph{{Finitely Reducible Realizations of the $N=2$ Superconformal Algebra}}, Nucl. Phys. B {\bf 313} (1989)
473--491

\bibitem{Fateev:1985mm}
V.~A. Fateev and A.~B. Zamolodchikov, \emph{{Parafermionic Currents in the Two-Dimensional Conformal Quantum Field Theory and Selfdual Critical Points in Z(n) Invariant Statistical Systems}}, Sov. Phys. JETP {\bf 62} (1985)
215--225

\bibitem{Ogawa:2024nhx}
N.~Ogawa, S.~Takahashi, T.~Tsuda  and T.~Waki, \emph{{Celestial CFT from $H_3^+$-WZW Model}},
\href{http://www.arXiv.org/abs/2404.12049}{{\tt 2404.12049}}

\bibitem{Mol:2024qct}
I.~Mol, \emph{{Partial Differential Equations for MHV Celestial Amplitudes in Liouville Theory}},
\href{http://www.arXiv.org/abs/2409.05936}{{\tt 2409.05936}}

\bibitem{Mol:2024onu}
I.~Mol, \emph{{Comments on Celestial CFT and $AdS_{3}$ String Theory}},
\href{http://www.arXiv.org/abs/2410.02620}{{\tt 2410.02620}}

\bibitem{Mol:2024vok}
I.~Mol, \emph{{An $AdS_{3}$ Dual for Supersymmetric MHV Celestial Amplitudes}},
\href{http://www.arXiv.org/abs/2411.14311}{{\tt 2411.14311}}

\bibitem{Bu:2022dis}
W.~Bu and E.~Casali, \emph{{The 4d/2d correspondence in twistor space and holomorphic Wilson lines}}, JHEP {\bf 11} (2022) 076,
\href{http://www.arXiv.org/abs/2208.06334}{{\tt 2208.06334}}

\bibitem{Witten}
E.~Witten, \emph{{On string theory and black holes}}, Phys. Rev. D {\bf 44} (1991)
314--324

\bibitem{DVV}
R.~Dijkgraaf, H.~L. Verlinde  and E.~P. Verlinde, \emph{{String propagation in a black hole geometry}}, Nucl. Phys. B {\bf 371} (1992)
269--314

\bibitem{KB}
M.~Bershadsky and D.~Kutasov, \emph{{Comment on gauged WZW theory}}, Phys. Lett. B {\bf 266} (1991)
345--352

\bibitem{Becker}
K.~Becker, \emph{{Strings, black holes and conformal field theory}}, phd thesis, Theory Division, CERN and Physikalisches Institut, Universität Bonn, 2, 1994.

\bibitem{Wakimoto:1986gf}
M.~Wakimoto, \emph{{Fock representations of the affine lie algebra A$_1^{(1)}$}}, Commun. Math. Phys. {\bf 104} (1986)
605--609

\bibitem{Ribault:2005ms}
S.~Ribault, \emph{{Knizhnik-Zamolodchikov equations and spectral flow in AdS$_3$ string theory}}, JHEP {\bf 09} (2005) 045,
\href{http://www.arXiv.org/abs/hep-th/0507114}{{\tt hep-th/0507114}}

\bibitem{GN3}
G.~Giribet and C.~A. Nunez, \emph{{Correlators in AdS$_3$ string theory}}, JHEP {\bf 06} (2001) 010,
\href{http://www.arXiv.org/abs/hep-th/0105200}{{\tt hep-th/0105200}}

\bibitem{Maldacena2001}
J.~M. Maldacena and H.~Ooguri, \emph{{Strings in AdS$_3$ and SL(2,R) WZW model 1: The Spectrum}}, J. Math. Phys. {\bf 42} (2001) 2929--2960,
\href{http://www.arXiv.org/abs/hep-th/0001053}{{\tt hep-th/0001053}}

\bibitem{Ribault:2005wp}
S.~Ribault and J.~Teschner, \emph{{$H^+_3$-WZNW correlators from Liouville theory}}, JHEP {\bf 06} (2005) 014,
\href{http://www.arXiv.org/abs/hep-th/0502048}{{\tt hep-th/0502048}}

\bibitem{Stoyanovsky}
A.~V. Stoyanovsky, \emph{{A relation between the Knizhnik-Zamolodchikov and Belavin-Polyakov-Zamolodchikov systems of partial differential equations}},
\href{http://www.arXiv.org/abs/math-ph/0012013}{{\tt math-ph/0012013}}

\bibitem{Hikida:2007tq}
Y.~Hikida and V.~Schomerus, \emph{{$H^+_3$-WZNW model from Liouville field theory}}, JHEP {\bf 10} (2007) 064,
\href{http://www.arXiv.org/abs/0706.1030}{{\tt 0706.1030}}

\bibitem{Giribet}
G.~Giribet, \emph{{Violating the string winding number maximally in Anti-de Sitter space}}, Phys. Rev. D {\bf 84} (2011) 024045, \href{http://www.arXiv.org/abs/1106.4191}{{\tt 1106.4191}},
[Addendum: Phys.Rev.D 96, 024024 (2017)]

\bibitem{DO}
H.~Dorn and H.~J. Otto, \emph{{Two and three point functions in Liouville theory}}, Nucl. Phys. B {\bf 429} (1994) 375--388,
\href{http://www.arXiv.org/abs/hep-th/9403141}{{\tt hep-th/9403141}}

\bibitem{ZZ}
A.~B. Zamolodchikov and A.~B. Zamolodchikov, \emph{{Structure constants and conformal bootstrap in Liouville field theory}}, Nucl. Phys. B {\bf 477} (1996) 577--605,
\href{http://www.arXiv.org/abs/hep-th/9506136}{{\tt hep-th/9506136}}

\bibitem{Giribet:2005ix}
G.~Giribet and Y.~Nakayama, \emph{{The Stoyanovsky-Ribault-Teschner map and string scattering amplitudes}}, Int. J. Mod. Phys. A {\bf 21} (2006) 4003--4034,
\href{http://www.arXiv.org/abs/hep-th/0505203}{{\tt hep-th/0505203}}

\bibitem{GN2}
G.~Giribet and C.~A. Nunez, \emph{{Aspects of the free field description of string theory on AdS$_3$}}, JHEP {\bf 06} (2000) 033,
\href{http://www.arXiv.org/abs/hep-th/0006070}{{\tt hep-th/0006070}}

\end{thebibliography}\endgroup

\end{document}